\documentclass[12pt,preprint]{aastex}

\shorttitle{A Region void of Jovian Irregulars}
\shortauthors{Haghighipour and Jewitt}

\begin{document}

\title{A Region Void of Irregular Satellites Around Jupiter}

\author{N. Haghighipour and D. Jewitt}
\affil{Institute for Astronomy and NASA Astrobiology Institute,\\
University of Hawaii, Honolulu, HI 96822}
\email{nader@ifa.hawaii.edu, jewitt@ifa.hawaii.edu}

\begin{abstract}
An interesting feature of the giant planets of our solar system is
the existence of regions around these objects where no irregular 
satellites are observed. Surveys have shown that, around Jupiter,
such a region extends from the outermost regular satellite Callisto, 
to the vicinity of Themisto, the innermost irregular satellite. 
To understand the reason for the existence of such a 
{\it satellite-void} region, we have studied the dynamical evolution 
of Jovian irregulars by numerically integrating the orbits of several 
hundred test particles, distributed in a region between 30 and 
80 Jupiter-radii, for different values of their semimajor axes, 
orbital eccentricities, and inclinations. As expected, our simulations
indicate that objects in or close to the influence zones of the
Galilean satellites become unstable because of interactions with 
Ganymede and Callisto. However, these perturbations 
cannot account for the lack of irregular satellites in the entire
region between Callisto and Themisto. It is suggested that at 
distances between 60 and 80 Jupiter-radii, Ganymede and Callisto 
may have long-term perturbative effects, which may require the 
integrations to be extended to times much longer than 10 Myr. 
The interactions of irregular satellites with protosatellites 
of Jupiter at the time of the formation of Jovian regulars 
may also be a destabilizing mechanism in this region. 
We present the results of our numerical simulations and discuss 
their applicability to similar satellite void-regions around other 
giant planets.
\end{abstract}

\keywords{planets and satellites: general, celestial mechanics,
solar system: general, methods: N-body simulations}

\section{Introduction}

Despite the differences in their compositions, structures and
mechanisms of formation, the giant planets of our solar system 
have one common feature. They all host irregular satellites. 
Thanks to wide field charge-coupled-devices (CCDs), the past 
few years have witnessed the discovery of a large number of 
these objects [see \citet{Jewitt07} for a comprehensive review]. 
At the time of writing of this article, 108 irregular satellites 
have been discovered, of which 55 belong to Jupiter, making 
the Jovian satellite system the largest among all planets.

Due to its proximity, the irregular satellites of Jupiter have
been the subject of extensive observational and theoretical 
research. Many of the dynamical characteristics of these objects, 
such as their orbital stability, dynamical grouping and their 
collision probability have long been studied 
\citep{Saha93,Carruba02,Nesvorny03,Nesvorny04,Beauge06,
Beauge07,Douskos07}. 
There is, however, one interesting feature in the distribution 
of Jovian irregulars that has not yet been fully understood. 
As shown by \citet{Sheppard03}, the region extending from the 
orbit of Callisto, the outermost Galilean satellite at 26 Jupiter-radii
$({R_J})$, to the periastron of Themisto $(\sim 76{R_J})$, Jupiter's
innermost irregular satellite, is void of irregulars. 

Observations suggest the presence of similar void
regions around all four giant planets. Table 1 and figure 1 show 
this in more detail.  As seen from figure 1, satellite void regions also exist 
{\it between} the
currently known irregular satellites of the giant planets. Theoretical
studies have indicated that there may be two possible scenarios for 
the existence of such void regions; 
ejection from the system due to mutual interactions with other
irregular satellites and, in the case of satellites that are
the remnants of collisions, clustering around their parent bodies
\citep{Kuiper56,Pollack79,Kessler81,Thomas91,Krivov02,
Nesvorny03,Nesvorny04,Beauge07}.
The focus of this paper is, however, on the lack of irregular 
satellites in the {\it boundary} between regulars and irregulars.
We are interested in understanding of why no irregular satellite exists
between the outermost Galilean satellite and Jupiter's innermost 
irregular one.

The lack of irregular satellites in the boundary between
regulars and irregulars may be attributed to the 
distribution of the orbits of the latter bodies. Since irregular satellites
appear to have been captured from heliocentric orbits, it may be
natural to expect them to preferably have large semimajor axes,
and therefore not to exist in close orbits.
Proving this to be so would be an important contribution to the 
subject, but, unfortunately,
none of the models of capture is sufficiently specific to be 
used in this way. The N-body capture model of \citet{Nesvorny07}
does roughly match the distribution of irregular 
satellites of some planets, but not of Jupiter.
In this paper, we examine the 
possibility of a dynamical origin for the existence of this 
satellite-void boundary region.

The origin of irregular satellites and the mechanisms of their
capture remain unknown. The high values of the orbital inclinations 
and eccentricities of these objects imply an origin outside the
primordial circumplanetary disk from which the regular satellites of
giant planets were formed. It is believed that irregular satellites 
were formed elsewhere and were captured in their current orbits 
\citep{Kuiper56,Pollack79,Nesvorny03,Nesvorny04,Jewitt07}. 

The capture of irregular satellites might have occurred during 
and/or after the formation of the regular satellites of the giant 
planets. Given that the latter objects are formed through the 
collisional growth of small bodies in a circumplanetary disk
\citep{Canup02,Mosqueira03a,Mosqueira03b,Estrada06},  
the orbits of captured irregulars might have been altered by 
perturbations from these objects during their formation and after 
they are fully formed. In the case of Jovian irregulars, the 
migrations of Ganymede and Callisto 
\citep{Tittemore88,Tittemore89,Tittemore90,Goldreich80,Canup02}
have also had significant effects on the dynamics of irregular 
satellites. 

In this paper we study the dynamics and stability of irregular satellites
between Callisto and Themisto. We present the details of our
model in \S 2, and an analysis of the results in \S 3. Section 4
concludes this study by reviewing our study and discussing its
limitations.

\section{Numerical Simulations}
We numerically integrated the orbits of 
several hundred test particles in a region interior to the orbit 
of Themisto, the innermost Jovian irregular satellite. We assumed 
that the regular satellites of Jupiter were fully formed and studied 
the perturbative effects of the Galilean satellites on the dynamics 
of small objects in their vicinities.
We considered a system consisting of Jupiter, the Galilean satellites, 
and 500 test particles uniformly distributed between 30 and 
80 Jupiter-radii. The initial orbital elements of the test 
particles were chosen in a systematic way as explained below.

\noindent
1) At the beginning of each simulation, test particles were placed
in orbits with semimajor axes starting at $30{R_J}$ and increasing 
in increments of 0.1$R_J$. 

\noindent
2) For each initial value of the semimajor axis of a test 
particle $(a_p)$, the initial orbital eccentricity $(e_p)$ 
was chosen to be 0, 0.2, 0.4, and 0.6. This choice of orbital 
eccentricity matches the range of the current values of the orbital
eccentricities of Jovian irregulars, as shown in figure 1. 

\noindent
3) The initial orbital inclinations of test particles $(i_p)$ 
were varied between $0^\circ$ and $180^\circ$ in steps of 
$20^\circ$. As shown by figure 2, irregular satellites are 
absent at inclinations between $55^\circ$ and $130^\circ$ due 
to perturbations resulting from the Kozai resonance
\citep{Kozai62,Hamilton91,Carruba02,Nesvorny03}.
In choosing the initial orbital inclinations of these objects, 
we made a conservative assumption and considered the region of 
the influence of Kozai resonance to be between $60^\circ$ and 
$120^\circ$. We did not integrate the orbits of the test particles 
for ${i_p}={80^\circ}, {100^\circ}$ and $120^\circ$.

\noindent
4) Since we were interested in studying the effects of the perturbations 
of regular satellites on the variations of the orbital eccentricities
and inclinations of test particles, we considered the initial values 
of the argument of the periastron, longitude of the ascending 
node, and the mean-anomaly of each test particle to be zero. 
This is an assumption that was made solely for the 
purpose of minimizing the initial-value effects.

We numerically integrated the orbits of the Galilean 
satellites\footnote{The orbital elements of the Galilean satellites 
were obtained from documentation on solar system dynamics 
published by the Jet Propulsion Laboratory 
(http://ssd.jpl.nasa.gov/?sat\_elem).} and the test particles 
of our system for different values of the test particles' 
orbital eccentricities and inclinations. Simulations were 
carried out for 10 Myr using the N-body integration package 
MERCURY \citep{Chambers99}. Since the objects of our interest 
are close to Jupiter, we neglected the perturbation of the Sun and 
considered Jupiter to be the central massive object of the system. 
This assumption is consistent with the findings of \citet{Hamilton97},
who have shown that around a giant planet with a Hill radius $R_H$,
the gravitational force of the Sun destabilizes the orbit of 
prograde irregular satellite at distances larger than 0.53$R_H$ 
and those of the retrograde irregulars at distances beyond 
0.69$R_H$. For Jupiter, these values translate to 389$R_J$ 
for prograde irregulars and 507$R_J$ for retrograde ones. 
We carried out all integrations with respect to Jupiter with 
timesteps equal to the 1/20 of the orbital period of Io.

\section{Discussion and Analysis of the Results}

To study the relation between the orbital parameters of test 
particles and their stability, we determined the lifetime of 
each particle, considering ejection from the system and 
collision with other bodies. We considered a particle to be 
ejected when it reached a distance of 2000$R_J$ or larger from 
the center of Jupiter. A collision, on the other hand, occurred 
when the distance between a particle and a Galilean satellite 
became smaller than ${R_{GH}}={a_{GS}}{({M_{GS}}/3{M_J})^{1/3}}$, 
or the particle's closest distance to the center of Jupiter 
became smaller than Jupiter's radius. Here, $a_{GS}$ and $M_{GS}$ 
represent the semimajor axis and mass of a Galilean satellite, and
$M_J$ is the mass of Jupiter. Figure 3 shows the graphs of the 
test particles' lifetimes in terms of their initial semimajor 
axes for particles in two coplanar systems. The graph at the 
top corresponds to particles initially in circular orbits, 
and the one at the bottom shows the lifetimes of particles 
with initial eccentricities of 0.2. The positions and lifetimes 
of the regular satellites of Jupiter and the orbit of Themisto 
are also shown. As shown by the upper graph, test particles in 
circular orbits are mostly stable (for the duration of integrations)
except for a few that are close to Callisto. The region of 
stability, however, becomes smaller (instability progresses 
toward larger distances) in simulations in which the initial 
eccentricities of test particles are larger. This can be seen 
more clearly in figure 4 where from the graphs of figure 3,
only the regions between $30{R_J}$ and $80{R_J}$ are shown. The 
islands of instability, corresponding to mean-motion resonances 
with Callisto (indicated by the subscript C) and Ganymede 
(indicated by the subscript G) are also shown.

The migration of unstable regions to larger distances in systems
where test particles were initially in eccentric orbits was observed 
in all our simulations. Figure 5 shows another example of such a 
system. In this figure, the lifetimes of test particles with initial 
eccentricities of 0.4 and initial inclinations of 20$^\circ$ are 
shown. The unstable region extends to distances beyond their
corresponding regions in figures 3 and 4.

We also simulated the dynamics of test particles having
orbital inclinations larger than 90$^\circ$ (retrograde orbits).
As shown by figure 2, the number of irregular satellites is larger 
at these angles implying that retrograde orbits have longer lifetimes 
\citep{Hamilton97,Touma98,Nesvorny03}. Our simulations also show 
that retrograde orbits are more stable than their corresponding 
prograde ones. Figure 6 shows this for two sets of test particles. 
The particles in black correspond to a system in which ${e_p}=0.4$ 
and ${i_p}=40^\circ$. The particles in red correspond to a system
with similar orbital eccentricity, but with ${i_p}=140^\circ$.
As expected, the particles on retrograde orbits are more
stable and maintain their orbits for longer times.

The fact that the region of instability of test particles,
having a given semimajor axis,  expands by increasing the initial 
values of their orbital eccentricities can be attributed to the 
interactions of these particles with Jupiter's regular satellites. 
Given that the orbits of Jovian regulars are almost circular, 
an eccentric orbit for a test particle implies a smaller periastron 
distance for this object, and consequently a closer approach to 
the system's regular moons. Instability occurs when the perturbative 
effects of regular satellites disturb the motion of a test particle 
in its close approach. The outer boundaries of the influence 
zones\footnote{We define the {\it influence zone} of a Galilean 
satellite as the region between $({a_{GS}}-3{R_{GH}})$ and 
$({a_{GS}}+3{R_{GH}})$, where the dynamics of a small object is 
primarily affected by the gravitational force of the satellite.} 
of the Galilean satellites (Table 2) mark the extent of these 
perturbations. Particles with periastron distances beyond these 
boundaries, i.e., ${a_p}(1-{e_p}) > ({a_{GS}}+3{R_{GH}})$, will 
more likely have longer lifetimes.  

Figure 7 shows the boundaries of the stable and unstable test 
particles for all Galilean satellites. The initial positions of 
test particles with $({e_p},{i_p})$ equal to (0,0), (0.2,0), 
(0.4,40), (0.4,140), (0.6,60), and (0.6,120), are also shown. 
The stable particles are shown in black and unstable ones are 
in red. As shown here, particles with higher initial orbital 
eccentricities penetrate the influence zones of Ganymede and 
Callisto, and their orbits become unstable. Figure 7 also shows 
that for particles with similar semimajor axes, the boundary 
between the stable and unstable regions extends to larger 
distances by increasing the particle's eccentricity. For instance, 
when interacting with Callisto, in order for a particle to 
maintain stability, ${a_p}(1-{e_p}) > 28 {R_J}$. However, for 
particles in figure 5, where ${e_p}=0.2$, this implies that the 
region of instability extends to at least $35{R_J}$. 

Although for a given semimajor axis, the boundary of stable and 
unstable regions expands with increasing orbital eccentricities 
of the test particles, for a given value of this eccentricity, the 
destabilizing effects of the Galilean satellites reach to larger
semimajor axis, beyond their influence zones. At such distances, 
although the perturbative effects of Galileans are small they may, 
in the long term, disturb the motions of other objects and render 
their orbits unstable. An example of such instability can be seen 
in figure 7 for $({e_p},{i_p})=(0.2,0), (0.4,40), (0.6,60)$ and 
also in figure 5, where the unstable region extends to approximately 
$46{R_J}$. These results also imply that in simulations similar 
to those shown in figure 3, the region of instability may migrate 
outward if the integrations are continued to much larger times 
beyond 10 Myr.

Ganymede and Callisto may have undergone inward radial migrations 
after their formation \citep{Goldreich80,Canup02}. As noted by 
\citet{Canup02}, Ganymede might have started its inward migration 
from a distance not larger than approximately $30{R_J}$, and 
Callisto might have migrated inward from approximately $35{R_J}$. 
The perturbative effects of these satellites have, therefore, 
influenced a larger region beyond their current influence zones. 
Figure 8 shows this in more detail. In this figure, the top graph 
shows the boundary of stable and unstable test particles for 
Ganymede before and after its radial migration. The bottom graph 
in figure 8 shows the similar curves for both Ganymede and Callisto. 
The influence zones of these two satellites were initially 
larger, implying that many small objects might have been destabilized 
during the migrations phase.

Even when including radial migration, the influence zones of 
Ganymede and Callisto do not cover the entire region between Callisto 
and Themisto. For the 10 Myr integration time presented here, the 
interactions with Ganymede and Callisto do not seem to account for 
the lack of irregular satellites at distances beyond 60$R_J$. Since 
at such distances, the perturbative effects of Ganymede and Callisto 
are weaker, extension of integrations to longer times may reveal that 
this region is indeed unstable. We also speculate that the lack of 
irregular satellites at such distances is the result of a clearing 
process that has occurred during the formation of the Jovian regular 
moons. As shown by \citet{Canup02}, and by 
\citet{Mosqueira03a,Mosqueira03b}, regular satellites of giant planets 
might have formed through the collisional growth of smaller 
objects (satellitesimals) in a circumplanetary disk. Similar to 
the formation of terrestrial planets in our solar system, where 
the mutual collisions of planetesimals around the Sun resulted in 
the formation of many protoplanetary objects, satellitesimals might 
have also collided and formed a disk of protosatellite bodies
around giant planets. The interaction between protosatellites and
smaller bodies in such circumplanetary disks could have destabilized
the orbits of many of these objects and resulted in their collisions,
accretion by protosatellites, and/or ejection from the system. 
The final locations of surviving irregular 
satellites at smaller distances were then limited by the outer 
boundary of a region that included the influence zones of regular 
satellites, as well as the above-mentioned dynamically cleared area.
For Jovian irregulars, a conservative assumption places this limit 
at $76{R_J}$ on the curve of constant-periastron of Themisto. 
Figure 9 shows this limit in light blue. At larger distances, on 
the other hand, the stability of irregular satellites
is governed by the perturbation from the Sun. The boundaries 
of the Sun-perturbed regions have been shown in figure 9 as curves 
of constant-apastron, with the constant value equal to 0.53$R_{JH}$ 
for prograde irregulars and 0.69$R_{JH}$ for retrograde ones 
\citep{Hamilton97}. The quantity $R_{JH}$
is the Hill radius of Jupiter. It is important to note that the values
of the eccentricities and semimajor axes of the irregular satellites
in figure 9 were obtained from the documentation on solar system dynamics
published by JPL, in which the orbital parameters of a body have their 
mean values and, unlike the test particles in our simulations, 
their angular elements are non-zero. This implies
that although figure 9 portrays a qualitatively reliable
picture of the stability of the Jovian irregular satellites, a
more detailed mapping could be obtained by assuming zero values for the
initial angular elements of irregular satellites and simulating
their stability for 10 Myr.

\section{Summary and Concluding Remarks}

We numerically integrated the orbits of 500 test particles 
for different values of their orbital elements, in a region 
between $30{R_J}$ and $80{R_J}$. Our integrations indicated 
that the long-term stability of these objects is affected 
by the values of their initial periastron distances. For given 
values of their semimajor axes, the region of instability of test 
particles extended to larger distances as the initial values 
of their orbital eccentricities were increased. 

Our numerical simulations also showed that, except at large
distances from the outer boundaries of the influence zones of 
Ganymede and Callisto, the lack of irregular satellites between 
Callisto and Themisto can be attributed to the instability of 
test particles caused by their interactions with the two outermost 
Galilean satellites. At larger distances (e.g., between 
$\sim 40{R_J}$ to 80$R_J$ for particles in circular orbits, 
and between $\sim 60{R_J}$ to 80$R_J$ for particles with initial 
orbital eccentricities of 0.4), however, the perturbations of 
Galilean satellites do not seem to be able to account for the 
instability of small bodies. A possible explanation is that their 
instability is the result of interactions with Jovian satellitesimals 
and protosatellites during the formation of Jupiter's regular moons.

Because the test particles in our simulations were initially close 
to Jupiter, we neglected the effect of solar perturbations.
As shown by \citet{Hamilton97}, for Jupiter, the shortest critical 
distance beyond which the perturbation from the Sun cannot be neglected
corresponds to prograde orbits and is equal to 389$R_J$.
In our simulations, the outermost test particle was placed
well inside this region at 80$R_J$. It is important to note that,
although the effects of solar perturbations
on our test particles are small and will not cause
orbital instability, they may, in the long term, create noticeable changes
in the orbital evolution of test particles. For instance, solar
perturbations may enhance the perturbative effects of regular satellites
in increasing the orbital eccentricity of test particles and
result in their capture in Kozai resonance. More numerical simulations 
are needed to explore these effects.

As mentioned in \S 2, different non-zero
angular elements may in fact affect the stability of individual 
test particles. However, the analysis of
the stability of the system, as obtained from our numerical 
simulations, portrays a picture of the dynamical
characteristics of the test particles that, in general, is also
applicable to Jovian irregular satellite systems with other 
initial angular variables.
 
The applicability of our results and the extension of our analysis 
to the satellite-void boundary regions around other giant planets may be 
limited due to fact that their satellite systems are different from 
that of Jupiter. Although the above-mentioned dynamical-clearing process
can still account for the instability of many small objects around 
these planets, numerical simulations, similar to those presented here, 
are necessary to understand the dynamical characteristics of their 
small bodies in more detail.

\acknowledgments
We acknowledge the use of the computational facilities
at the Department of Terrestrial Magnetism at the
Carnegie Institution of Washington. 
This work has been supported by the NASA Astrobiology
Institute under Cooperative Agreement NNA04CC08A at the Institute 
for Astronomy at the University of Hawaii for NH.

\clearpage
\begin{figure}
\plottwo{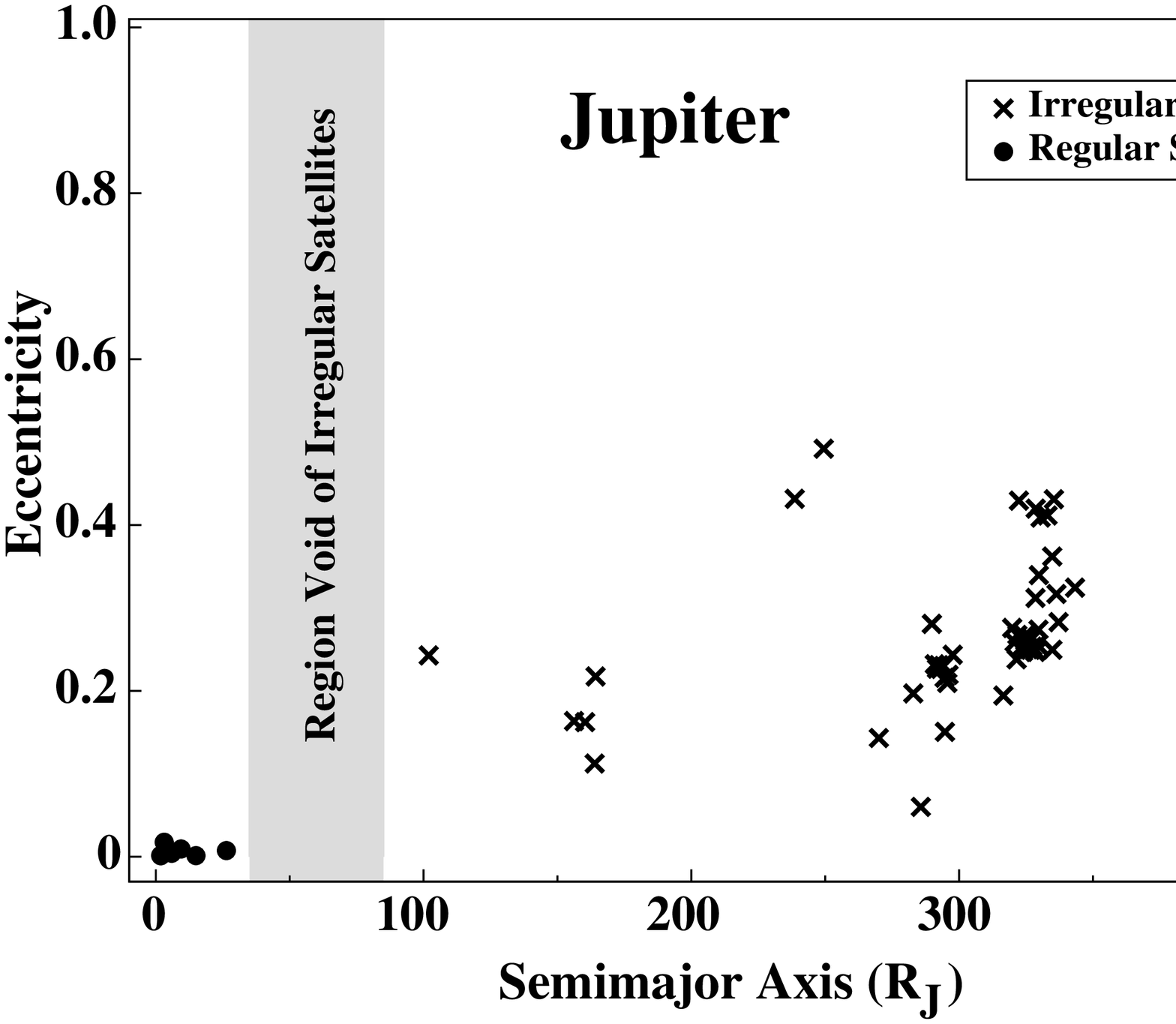}{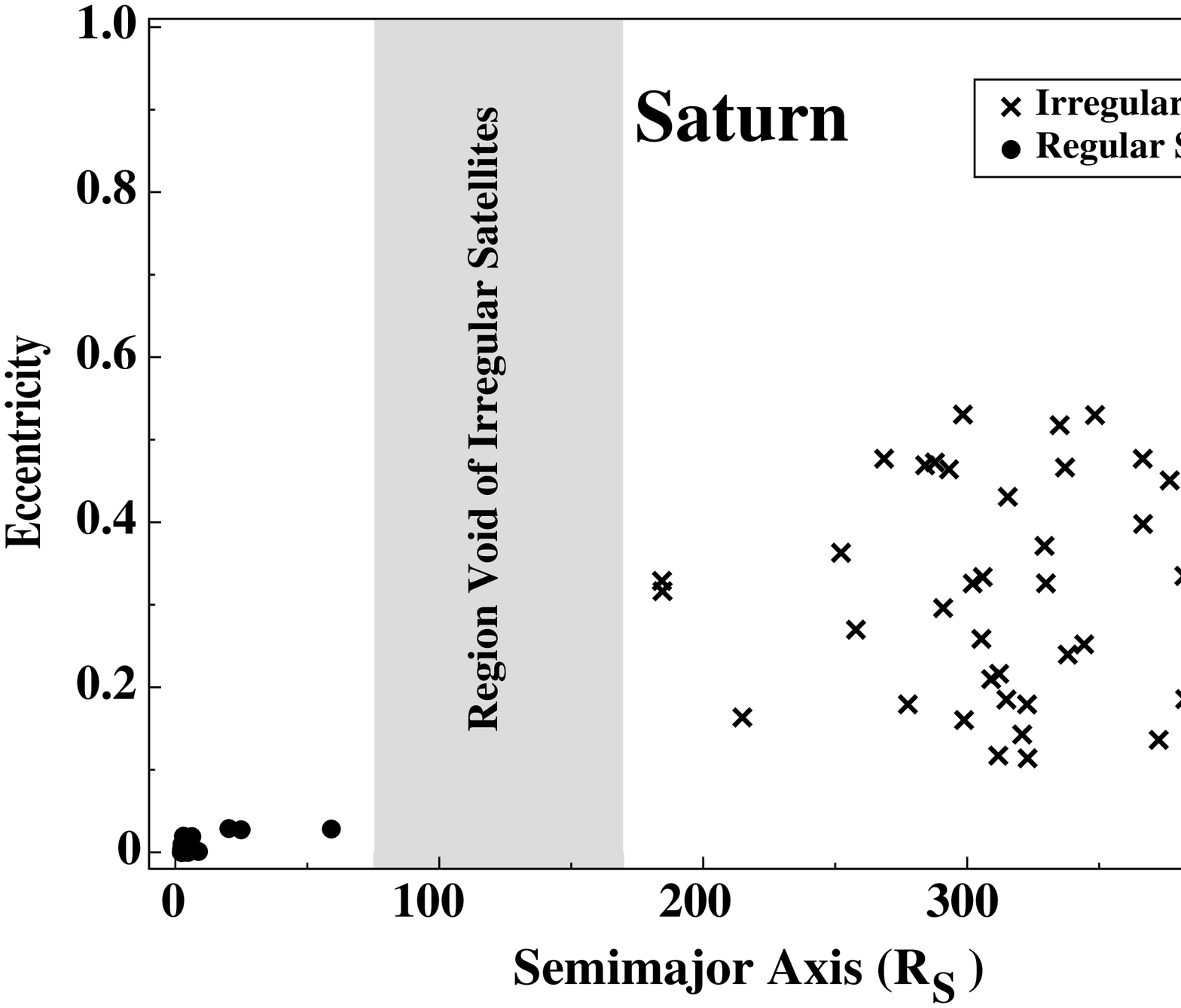}
\plotone{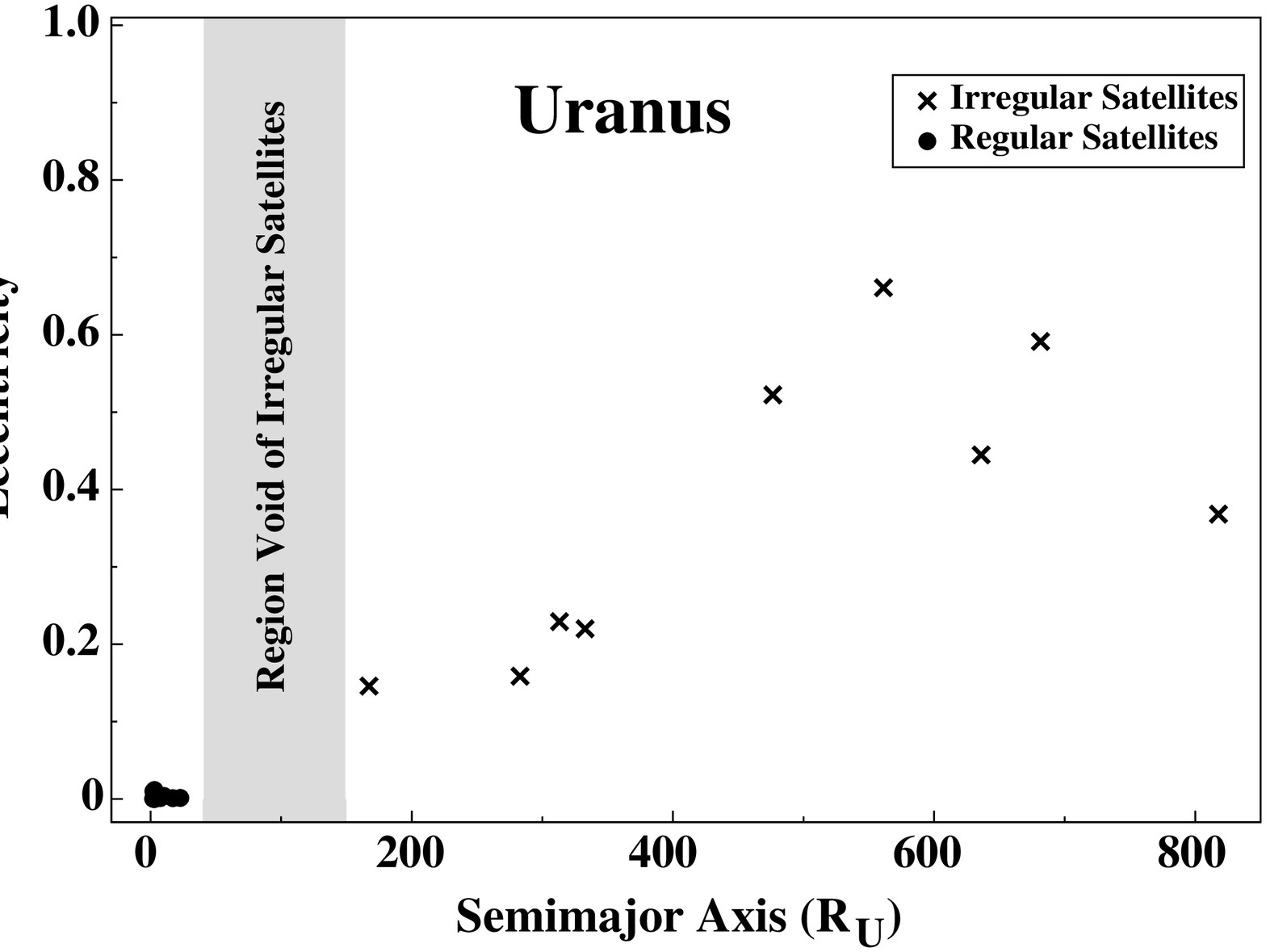}
\plotone{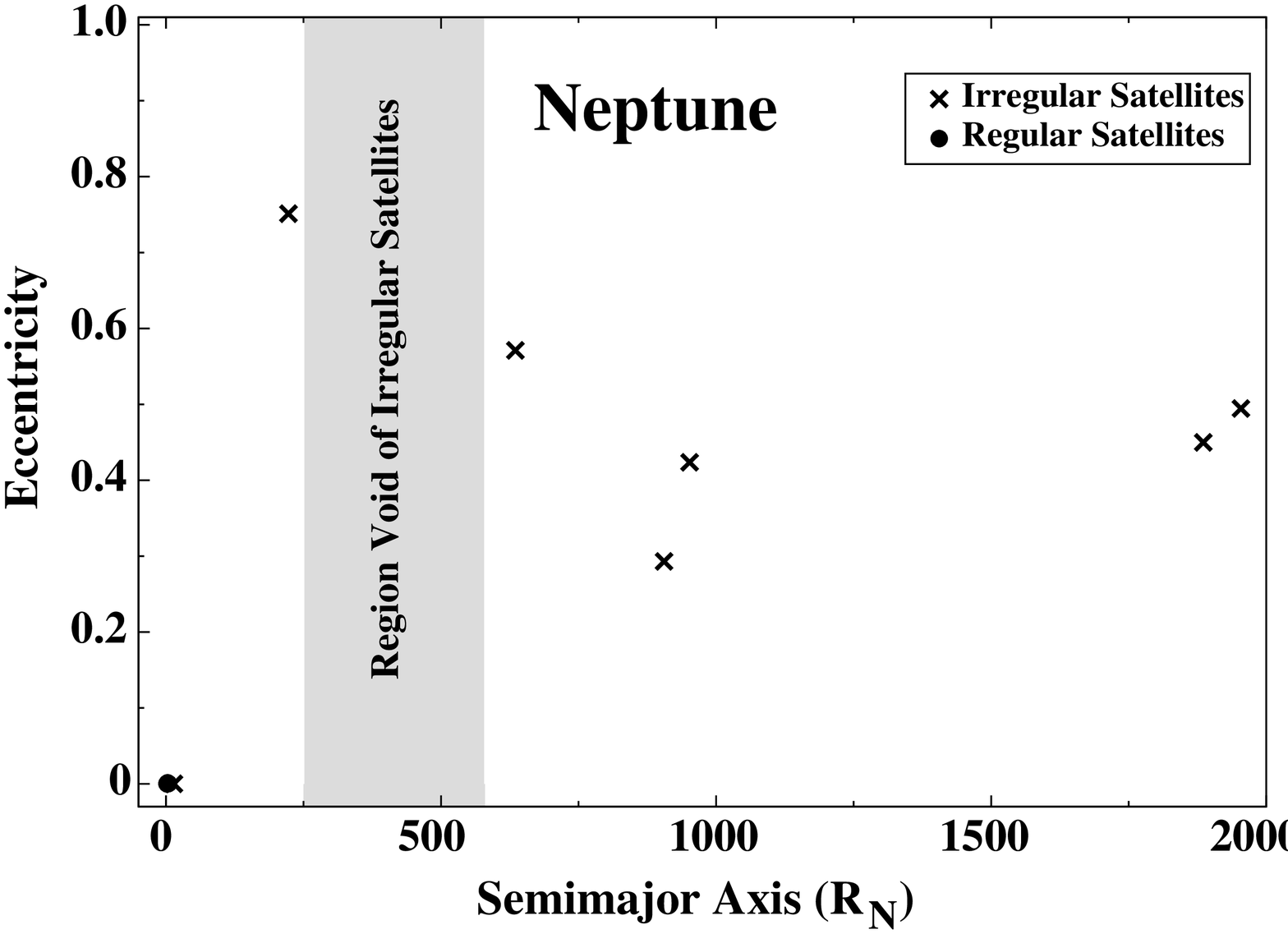}
\vskip 0.1in
\caption{Satellite systems of the four giant planets. The 
$x-$axes show the semimajor axes of satellites in the units
of the radii of their host planets.
\label{fig1}}
\end{figure}

\clearpage
\begin{figure}
\plotone{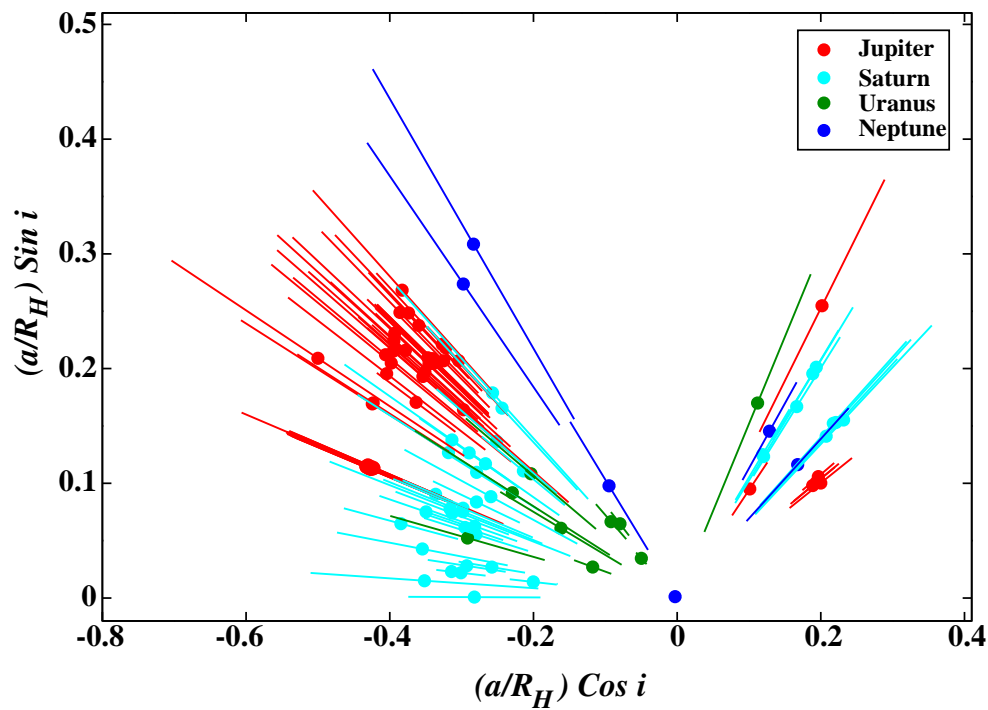}
\vskip -4.2in
\caption{Distribution of irregular satellites around the 
four giant planets. The quantities on the axes represent 
the semimajor axis of a satellite, $a$, the Hill radius 
of its host planet, $R_H$, and the satellite's orbital 
inclination, $i$. The distance of each satellite from the 
origin of the graph is equivalent to its semimajor axis, 
and its radial excursion (the distance from its periastron 
to apastron) is given by the length of its associated line. 
The angle between this line and the horizontal axis is 
equal to satellite's orbital inclination.
\label{fig2}}
\end{figure}

\clearpage
\begin{figure}
\vskip 2in
\includegraphics[height=8cm]{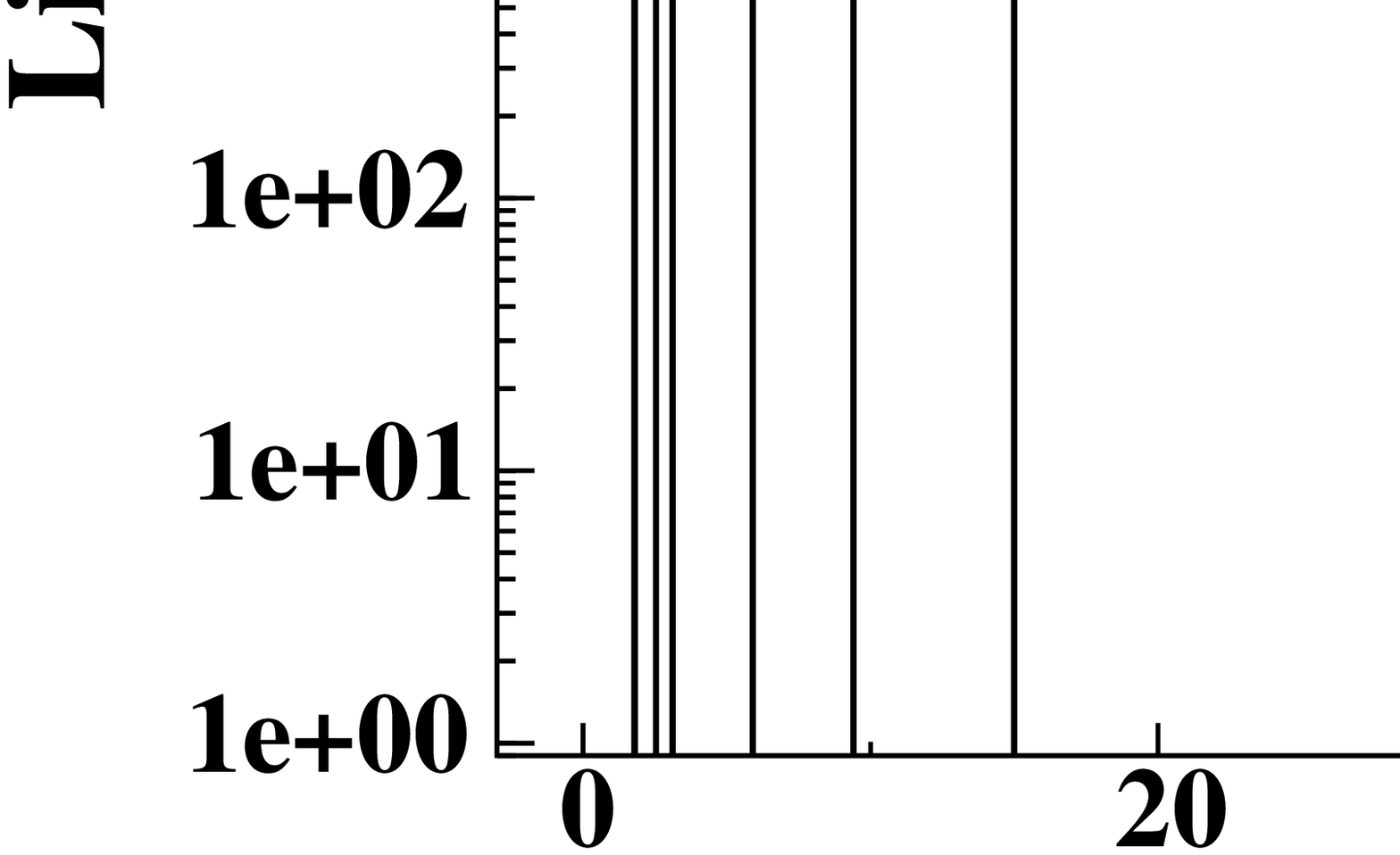}
\includegraphics[height=8cm]{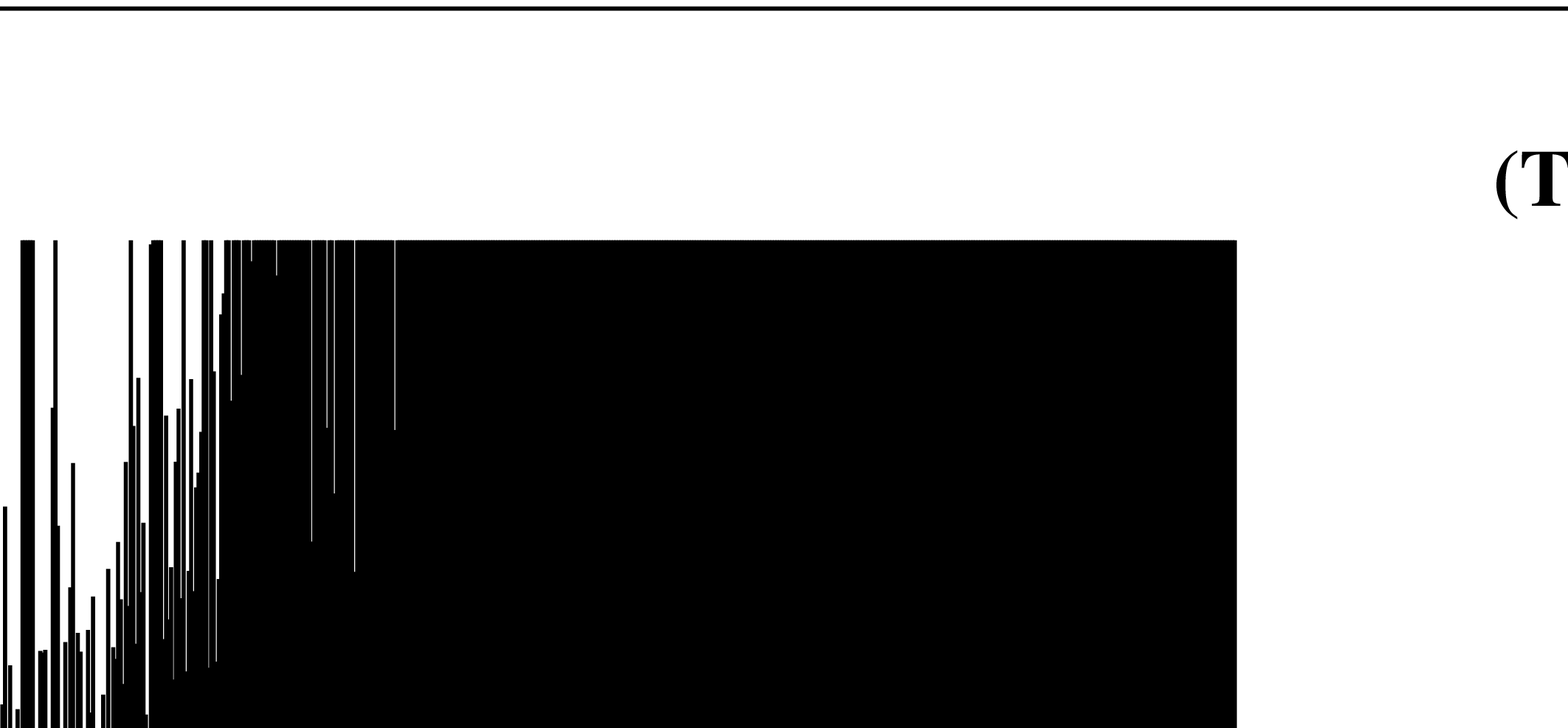}
\vskip 2.2in
\caption{Lifetimes of test particles in a coplanar system
with ${e_p}=0$ (top) and ${e_p}=0.2$ (bottom). The locations of
the regular satellites of Jupiter and Themisto are also shown.
\label{fig3}}
\end{figure}

\clearpage
\begin{figure}
\vskip 0.85in
\includegraphics[height=8cm]{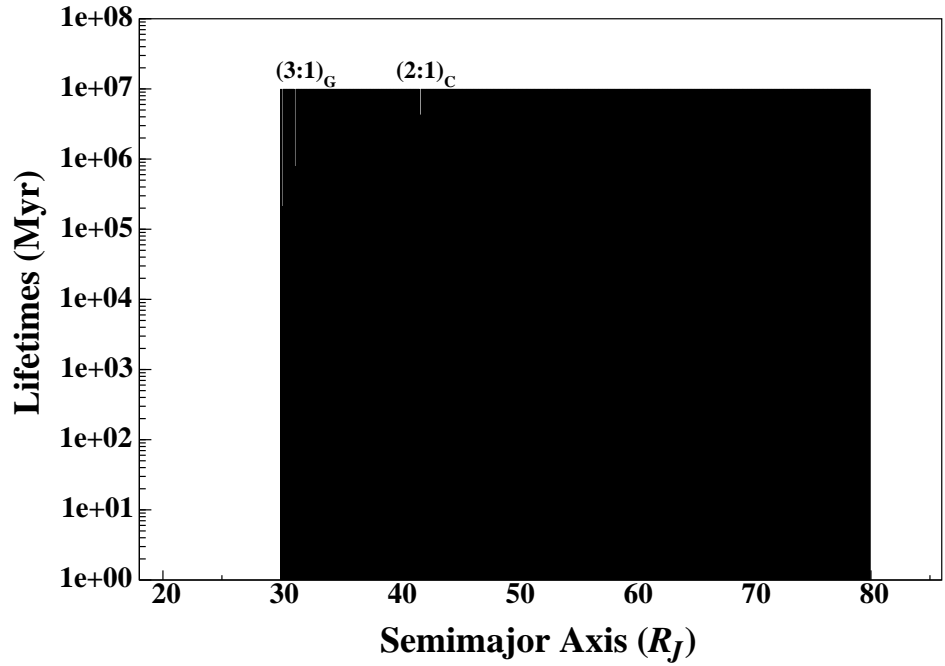}
\includegraphics[height=8cm]{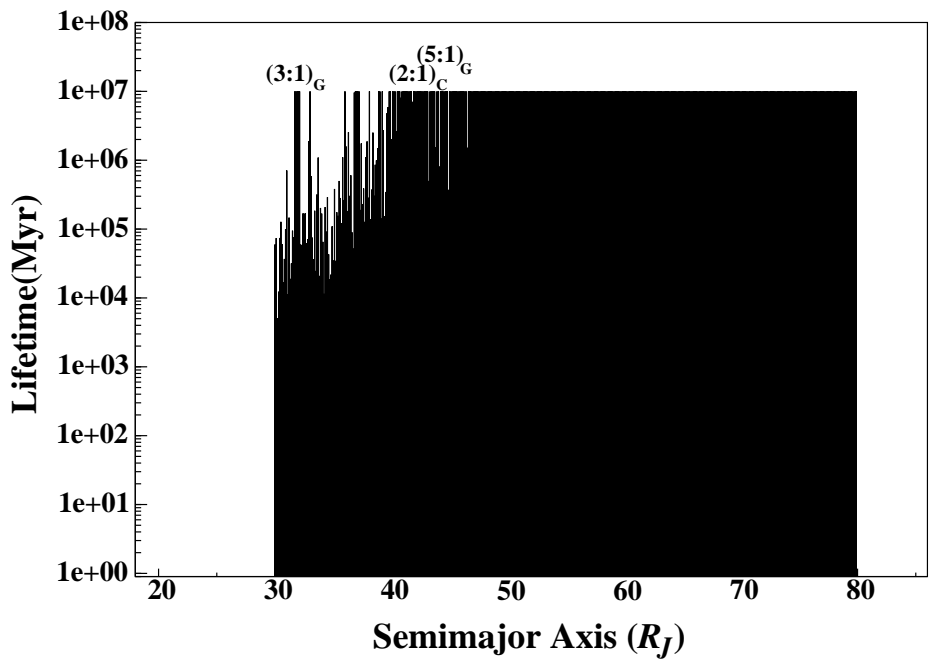}
\vskip 35pt
\caption{Lifetimes of the test particles of figure 3. Both 
systems are coplanar. In the upper graph ${e_p}=0$, and in 
the lower graph ${e_p}=0.2$. The locations of mean-motion 
resonances with Ganymede and Callisto are also shown.
\label{fig4}}
\end{figure}

\clearpage
\begin{figure}
\vskip 0.8in
\includegraphics[height=8cm]{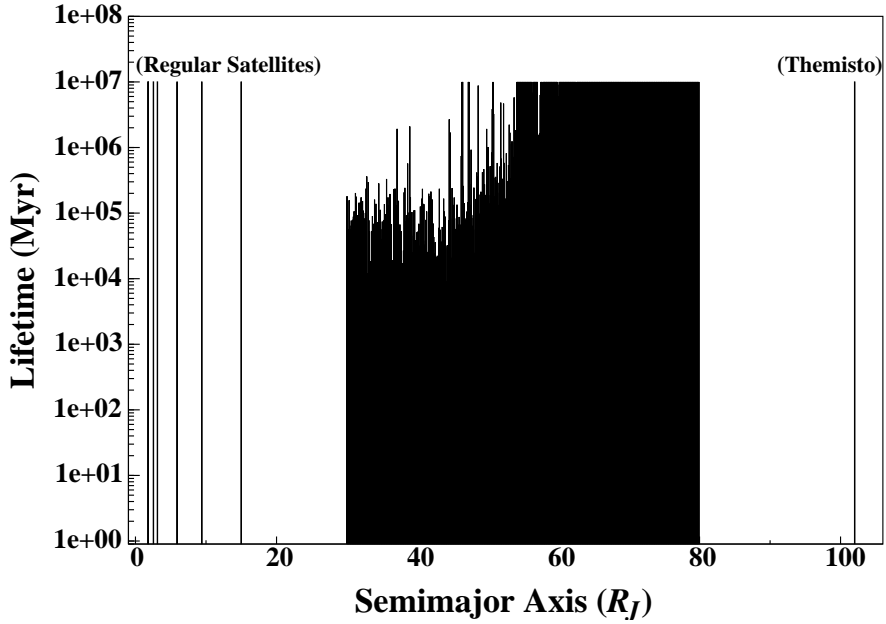}
\includegraphics[height=8cm]{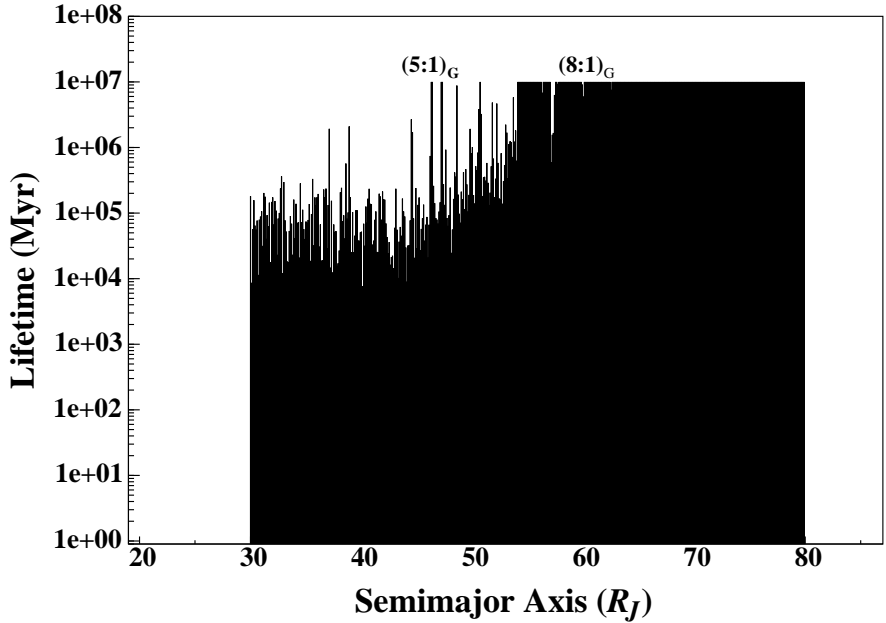}
\caption{Lifetimes of test particles in a system in which 
${e_p}=0.4$ and ${i_p}=20^\circ$. The locations of
mean-motion resonances with Ganymede and Callisto are also shown.
As shown here, compared with the systems of figures 3 and 4, 
as the initial eccentricities of test particle increase, their
region of instability extends to farther distances.
\label{fig5}}
\end{figure}

\clearpage
\begin{figure}
\centering
\includegraphics[height=9cm]{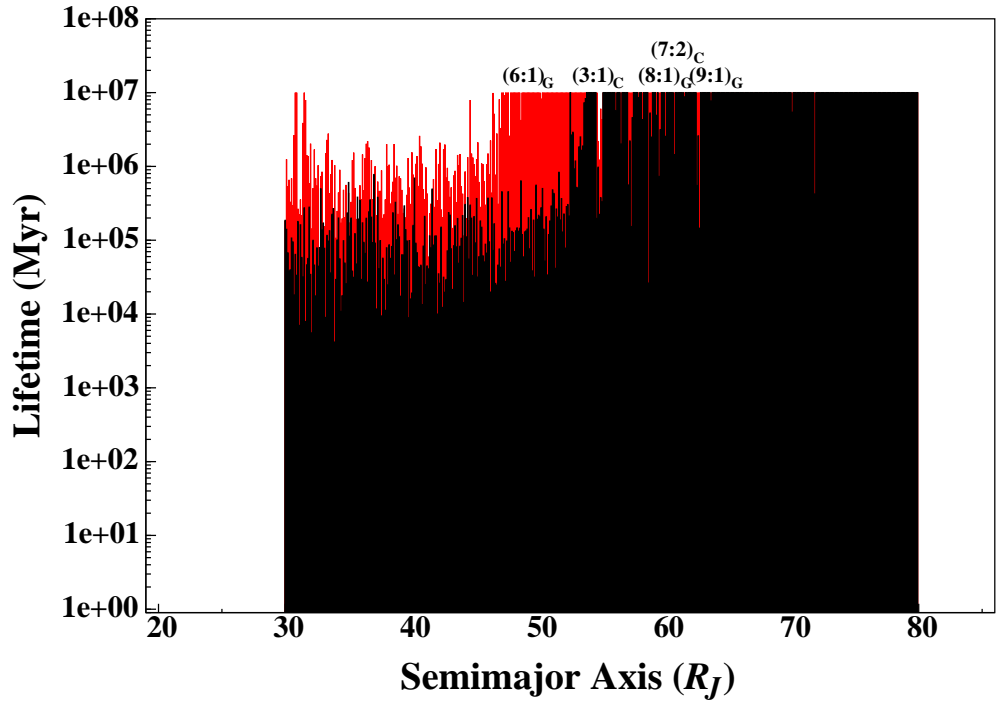}
\vskip 10pt
\caption{Lifetimes of test particles with initial eccentricities of
${e_p}=0.4$ and orbital inclinations of ${i_p}=40^\circ$ (black)
and ${i_p}=140^\circ$ (red). As shown here, particles in retrograde
orbits (red) are more stable. The locations of mean-motion resonances 
with Ganymede and Callisto are also shown. 
\label{fig6}}
\end{figure}

\clearpage
\begin{figure}
\centering
\includegraphics[height=10.5cm]{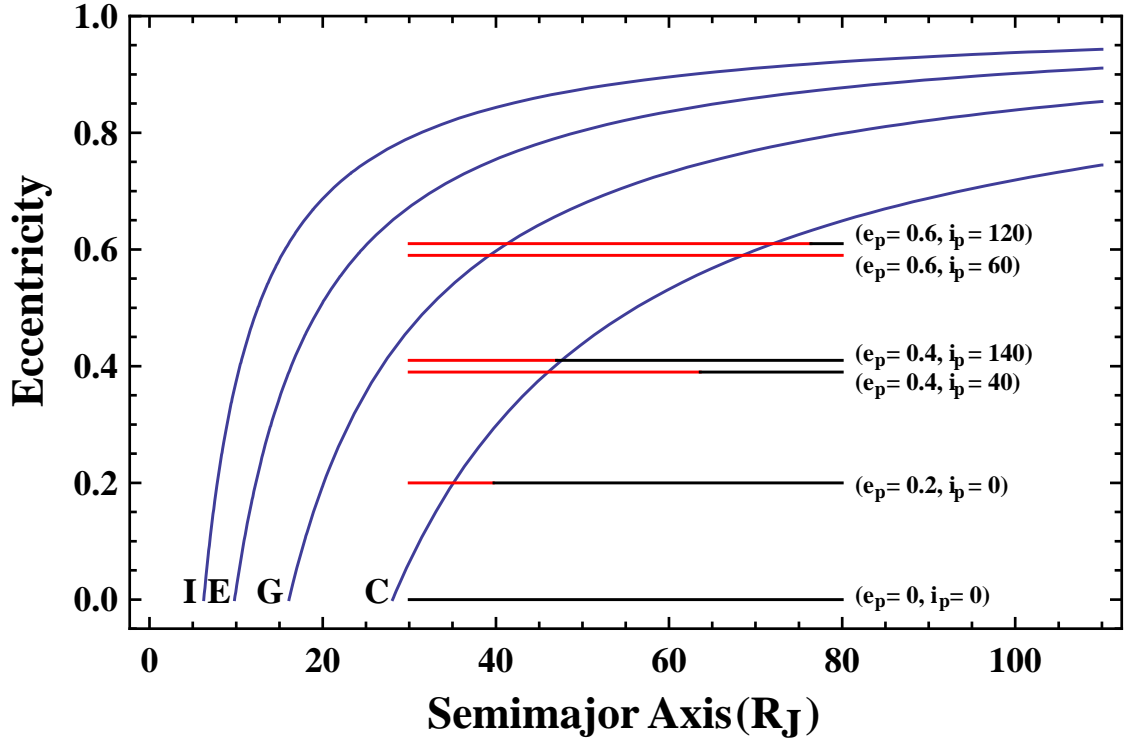}
\caption{Boundaries of stable and unstable regions of a test 
particle (i.e., curves of constant-periastron for which the 
constant value is equal to the distance of the outer 
boundary of the influence zones of the Galilean satellites). 
The black and red horizontal lines represent the initial semimajor
axes of test particles between $30{R_J}$ and $80{R_J}$ for
different values of their orbital eccentricities and inclinations.
The particles in black maintained their orbits for the duration of 
integration (10 Myr) whereas the particles in red became unstable.
\label{fig7}}
\end{figure}

\clearpage
\begin{figure}
\centering
\vskip -0.3in
\includegraphics[height=9cm]{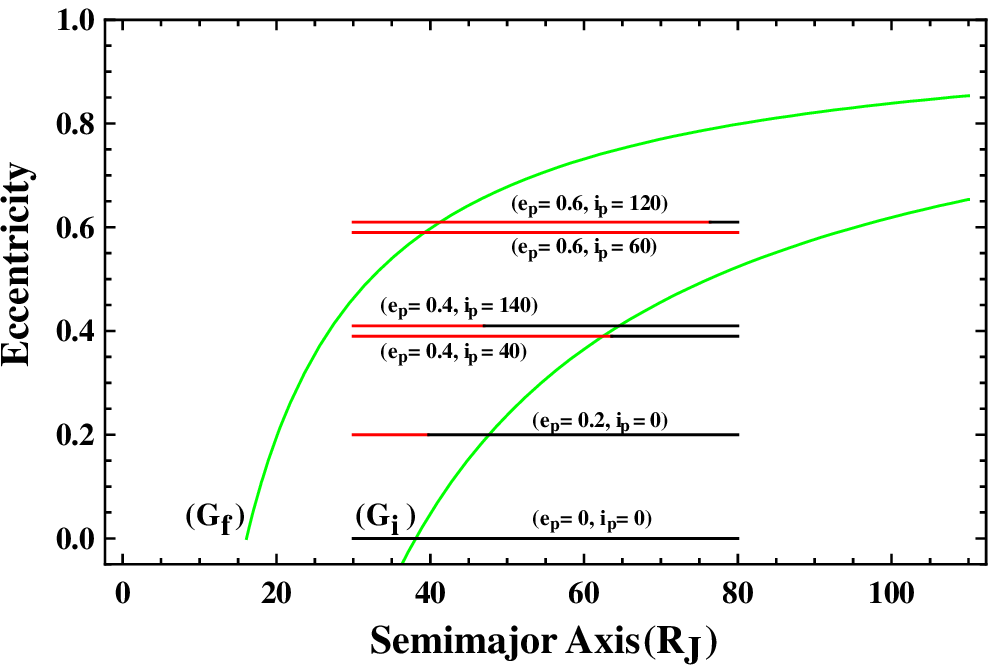}
\vskip 10pt
\includegraphics[height=9cm]{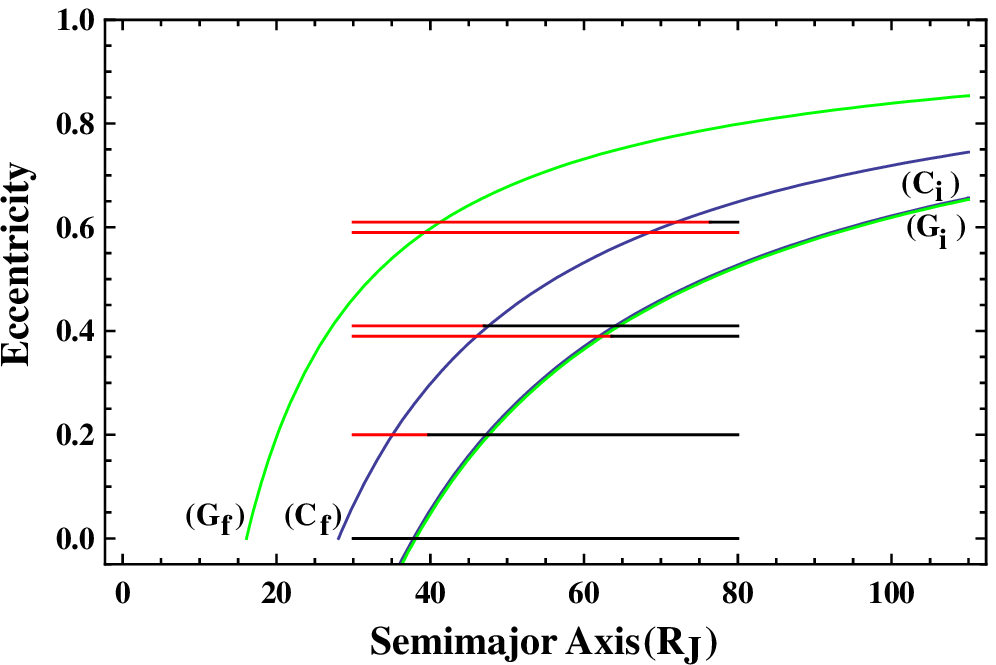}
\vskip -10pt
\caption{Top: curves of the constant-periastron (boundaries 
of stable and unstable regions) of a test particle for which 
the constant values are equal to the distance of the outer 
boundary of the influence zone of Ganymede before its migration
(the curve denoted by ${\rm G}_{\rm i}$) and after it migrates to 
its present orbit (the curve denoted by ${\rm G}_{\rm f}$).
The stable (black) and unstable (red) test particles in 
the region between $30{R_J}$ and $80{R_J}$ are also shown. 
Bottom: Similar constant-periastron curves as in the above
for Ganymede and Callisto. Note that, because the pre-migration
semimajor axes of Ganymede and Callisto are close to one
another ($30{R_J} $ and $35{R_J}$, respectively), it seems
as though the two
curves ${\rm G}_{\rm i}$ and ${\rm C}_{\rm i}$ in the bottom graph
are in contact with one another. The stable
and unstable test particle are also similar to the top graph.
\label{fig8}}
\end{figure}

\clearpage
\begin{figure}
\centering
\vskip 1.9in
\includegraphics[height=12cm]{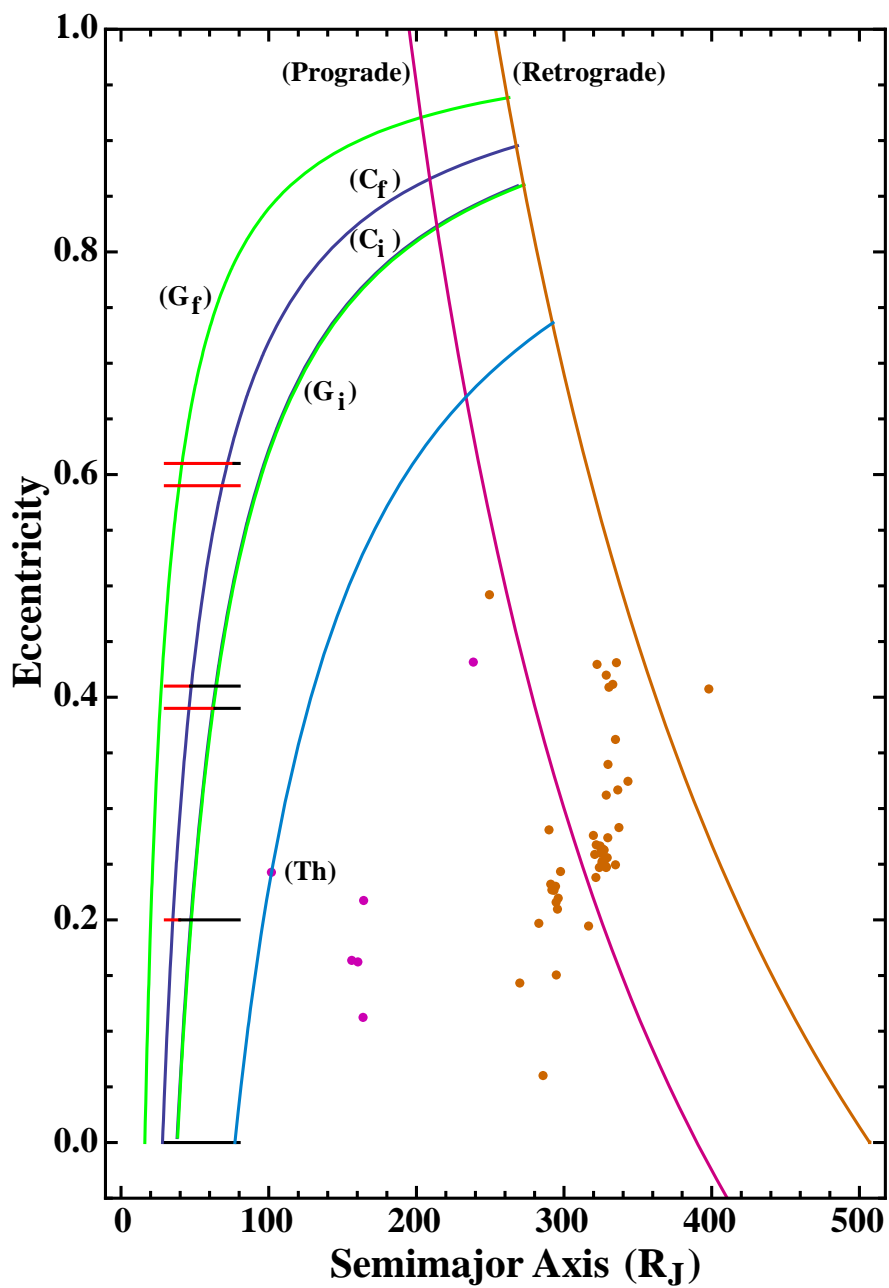}
\vskip -15pt
\caption{Graph of the region of the stability of Jovian 
irregular satellites. Prograde satellites are in purple 
and retrograde ones are in orange. The inner boundary of 
this region, shown in light blue, corresponds to the curve 
of constant-periastron of Themisto. Its outer boundary,
shown in purple for prograde irregulars and in orange for 
retrograde ones, is a curve for constant-apastron equal to 
the largest distance a Jovian irregular satellite can travel 
before its orbit becomes unstable by solar perturbation. The 
curves of constant-periastron corresponding to the influence 
zones of Ganymede and Callisto, as explained in figure 8, and 
the stable (black) and unstable (red) test particles, within 
the region of $30{R_J}$ to $80{R_J}$, are also shown.
\label{fig9}}
\end{figure}

\clearpage
\begin{deluxetable}{lc} 
\tablewidth{0pt} 
\tablecaption{Irregular satellite-void regions around giant planets}
\tablehead{ 
\colhead{Satellite} &
\colhead{Void Region (Planet Radii)}}
\startdata 
Jupiter   & 30-80         \\
Saturn    & 59-184         \\
Uranus    & 23-167         \\
Neptune   & 223-635           \\    

\enddata 
\end{deluxetable}

\clearpage
\begin{deluxetable}{lccc} 
\tablewidth{0pt} 
\tablecaption{Influence zones of the Galilean Satellites.}
\tablehead{ 
\colhead{Satellite} &
\colhead{Semimajor Axis $(R_J)$} &
\colhead{Inner boundary $(R_J)$} &
\colhead{Outer boundary $(R_J)$}}
\startdata 
Io         & 5.8     & 5.4      & 6.3         \\
Europa     & 9.3     & 8.7      & 9.8          \\
Ganymede   & 14.8    & 13.5     & 16.1          \\
Callisto   & 26.0    & 24.0     & 28.0           \\    

\enddata 
\end{deluxetable}

\end{document}